# Self-assembly of polymers or copolymers and ferrofluids leading to either 1-d, 2-d or 3-d aggregates decorated with magnetic nanoparticles

D. El kharrat[*], O. Sandre[*], R. Perzynski[†], F. Chécot[‡], S. Lecommandoux[‡]

[*]Laboratoire Liquides Ioniques et Interfaces Chargées
Unité Mixte de Recherche 7612 Centre National de la Recherche Scientifique / Université Pierre et Marie Curie
4, place Jussieu – case 63, 75252 Paris cedex 05 France
[†]Laboratoire Milieux Désordonnés et Hétérogènes
Unité Mixte de Recherche 7603 Centre National de la Recherche Scientifique / Université Pierre et Marie Curie
4, place Jussieu – case 78, 75252 Paris cedex 05 France
[‡]Laboratoire Chimie des Polymères Organiques
Unité Mixte de Recherche 5629 Centre National de la Recherche Scientifique / Ecole Nationale Supérieure de Chimie et de Physique de Bordeaux – 16, avenue Pey Berland, 33607 Peyssac France

A novel type of hybrid colloids is presented, based on the association of several polymeric systems and ferrofluids. On the one hand, we use inorganic nanoparticles made of magnetic iron oxide prepared at the LI2C, which response to a magnetic field of low intensity. On the other hand the organic part is made either of long linear polyacrylamide chains or of mesoscopic structures (vesicles and micelles) self-assembled from amphiphile polybutadiene-b-poly(glutamic acid) di-block copolymers, which conformation is pH-sensitive. Those PB-b-PGA copolymers bearing a cross-linkable hydrophobic block and a hydrophilic peptidic block have been synthesized recently at the LCPO by combining anionic polymerization and ring-opening polymerization[1]. Their polydispersity indices are small enough to obtain well defined self-assembled aggregates. For example the $PB_{40}$-b-$PGA_{100}$ copolymer leads in water to closed bilayers. Belonging to this new class of polymeric vesicles called polymersomes by F. Bates[2] (see also [3] for a review on polymer vesicles), they also exhibit the particularity of responding to a an external pH change by a significant variation of size: their hydrodynamic diameter $d_H$ varies indeed from 200nm in an acidic medium to 300nm in a basic medium[4] (or 250nm in concentrated brine). This transition with pH is fully reversible and only moderately sensitive to salinity for it is not based on a simple polyelectrolyte swelling effect but on the peptidic nature of the PGA block, which exhibits a real transition between a compact helix conformation at acidic pH and a stretched coil conformation at basic pH.

As for the magnetic nanoparticles, they come from ionic ferrofluids which are colloidal suspensions of nanometric magnetic grains stabilized either in aqueous media by electrostatic charges, or in organic solvents by appropriate tensioactives. Before grafting these dispersant molecules, the precursor ferrofluid [5] is prepared by alkaline co-precipitation of $FeCl_2$ and $FeCl_3$ leading to $Fe_3O_4$ (magnetite) nanocrystals, followed by surface sign reversal with $HNO_3$ and complete oxidation using $Fe(NO_3)_3$, which yields positively charged $\gamma$-$Fe_2O_3$ (maghemite) particles in $HNO_3$. One of the goals of embedding maghemite nanoparticles in the supra-macromolecular objects formed by di-block copolymers consists in utilizing the particles as probes for neutrons scattering in order to elucidate the strong morphological transformation of the vesicles during the size transition between acidic and basic pH. One expects indeed to detect a transient opening of the bilayer and/or a leak-out of the encapsulated content. For this purpose, we benefit from the very strong contrast of the iron oxide particles compared to polymer and solvent by measuring their structure factor inside the aggregates: this $S_{intra}(Q)$ gives indeed both the mean inter-particle distance[6] (hence their local concentration) and the global form factor of the object that they decorate[7]. Another guiding idea is to bring to the whole object a response to a magnetic field, either an induced shape change (ellipsoidal or not) or to trigger the delivery of an active substance by the application of a field.

This work started by the verification that those organic and inorganic systems can effectively be combined together to generate well defined objects which can be properly dispersed in a solvent. Therefore we examined a series of PB-b-PGA copolymers which differ by the length of their polypeptid PGA block, their hydrophobic block PB being chosen to enable an easy way to cross-linking within the frame of a future freezing of the structures. In this paper, we present a few cases representative of the reachable structures:
- $PB_{48}$-b-$PGA_{114}$ and $PB_{48}$-b-$PGA_{145}$ which self-assemble in water as micelles of hydrodynamic diameters $d_H$=60nm and $d_H$=70nm respectively, the internal diameter of the hydrophobic core being $d_{int}$=14nm as measured by SANS in both cases[8].
- $PB_{48}$-b-$PGA_{56}$ which forms closed membranes in water, i.e. vesicles characterized by an outer diameter $d_H$ =100nm and a bilayer thickness $\delta$=14nm[8].

We report also results obtained with a commercial polyacryamide homopolymer (Polysciences) known as to strongly interact with metal oxide colloids. This very long linear PAM 5–6x$10^6$g/mol is commonly used as flocculent. We tested the association of these two copolymers and this homopolymer with three either aqueous or organic ferrofluids, which grain size polydispersity had been reduced in a first step by the method of fractionated phase separations[9] with added excess $HNO_3$:
- $S1S$-$HNO_3$ acidic ferrofluid, which is an aqueous acidic ferrofluid (pH=1.2) with a particle size distribution characterized by a Log-normal law of parameters $d_0$=6.6nm and $\sigma$=0.21. Its iron oxide surface is coated simply by $H^+$ protons, thereby leaving a free access for a future grafting by the glutamic acid moieties of the copolymers or by the amide groups of polyacryamide.
- S1S-Na$_3$Cit citrated ferrofluid, which comes from the previous acidic ferrofluid after coating by tri-sodium citrate ligands, thus enabling a dispersion at pH=7.
- S2-CH$_2$Cl$_2$ surfaced ferrofluid, which is grafted by a phosphoric di-ester type tensioactive (Beycostat NE) for a solubilization in dichloromethane. It also has a slightly broader distribution of particle sizes ($d_0$=6.8 nm, $\sigma$=0.24 ).

Table 1. Short summary of samples and self-assembled hybrid objects including their dimensionality:

|  | PB$_{48}$-PGA$_{114}$ and PB$_{48}$-PGA$_{145}$ | PB$_{48}$-PGA$_{56}$ | PAM |
|---|---|---|---|
| S2-CH$_2$Cl$_2$ | (A) 3-d filled magnetic micelles | (B) 2-d soft magnetic shells in water | no interaction |
| S1S-HNO$_3$ | no proper dispersion | 3-d filled micelles and few 2-d vesicles in CH$_2$Cl$_2$ | 3-d hybrid network of fibers in water |
| S1S-Na$_3$Cit | no interaction | (C) 3-d filled magnetic vesicles | (D) 1-d necklaces |

The assembly of each copolymer mixed with either 1 or 2 equivalents of the surfaced ferrofluid S2-CH$_2$Cl$_2$ was favored by first eliminating the dichloromethane then redispersing the objects in water, after deprotonating the acidic groups of the copolymer with NaOH to raise the pH up to ≈7. In the case of the bare acidic particles S1S-HNO$_3$, the flocculation was induced almost instantaneously after addition of either PAM or of a PB-b-PGA copolymer. The supernatant could then be easily replaced by CH$_2$Cl$_2$. Finally with the S1S-Na$_3$Cit ferrofluid, the interaction with all the polymers did not lead to any visible destabilization.

After three weeks, all suspensions were analyzed by dynamic light scattering (DLS) and small angle neutrons scattering[10] (SANS). Thus we could show that some of the (co)polymer–particles aggregates reach an equilibrium shape and a proper dispersion state.

(A). Sample PB$_{48}$-PGA$_{145}$ mixed with the S2-CH$_2$Cl$_2$ surfaced ferrofluid leads to a dispersion of magnetic micelles in water, which diameters $d_H$ are equal to 430nm and 225nm for 1 and 2 mass equivalents respectively. The large increase of the outer diameter compared to unloaded micelles ($d_H$=70nm) suggest that their hydrophobic cores are filled with nanoparticles at a high local concentration. The short inter-particle distance given around 80Å by the position of the structure peak (Fig.1) corresponds to a volume fraction of nanoparticles inside the micelles about 30%. This high encapsulation yield together with the global 3-d shape of the aggregates are confirmed by a look at the TEM and AFM images in the case of PB$_{48}$-PGA$_{114}$, which show large and thick baggies of inorganic particles. Please note that the copolymer samples combined with 2 mass equivalents of ferrofluid lead to final volume fractions $\Phi$ of nanoparticles (from titration of iron) which are 6–7 times smaller than those with 1 mass equivalent. This discrepancy originates from the fact that the objects are fully dispersed only in the former case but not in the latter case. Thus at 2 equiv., we get supernatants which are globally more dilute in copolymer–particles complexes.

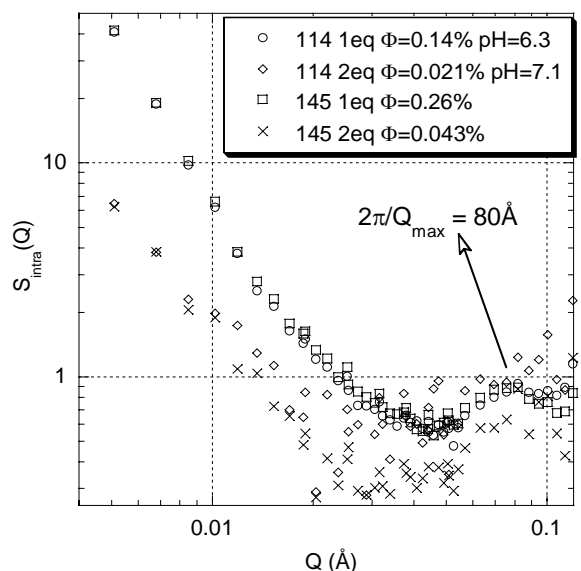

Fig.1. Intra-aggregates structure factor $S_{intra}(Q)$ for hydrophobic nanoparticles associated to PB$_{48}$-PGA$_{114}$ and PB$_{48}$-PGA$_{145}$, as measured by SANS from the scattered intensity of the particle–copolymer complexes normalized by the volume fractions $\Phi$ of particles and divided by the form factor $P(Q)$ of the S2-CH$_2$Cl$_2$ ferrofluid alone.

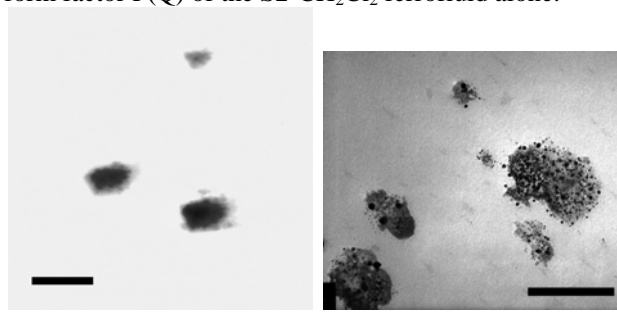

Fig. 2 AFM (left) and TEM (right) pictures of hybrid copolymer micelles loaded with magnetic nanoparticles, obtained by combining PB$_{48}$-PGA$_{114}$ and 1 mass equiv. of surfaced ferrofluid S2-CH$_2$Cl$_2$. The scale bar measures 400nm in both cases. The hydrodynamic diameter vary from $d_H$=60nm for the unloaded micelles to $d_H$=340nm and $d_H$=260nm for micelles loaded with 1 and 2 equivalents of hydrophobic particles respectively.

(B). Vesicles of PB$_{48}$-PGA$_{56}$ still form in water when in the presence of S2-CH$_2$Cl$_2$, their outer diameter being significantly increased ($d_H$ equal to 620nm and 210nm for 1 and 2 mass equivalents respectively). Due to their hydrophobicity, the surfaced nanoparticles are confined in

2-d between the two leaflets of the copolymer bilayer, as proved separately by the SANS measurements, the TEM pictures and the AFM imaging (which in addition contains a valuable topographical information). To the authors' knowledge this is the first described case of vesicles with a fluid magnetic membrane. Several magnetic shells were already mentioned in literature[11,12], but none as hollow and floppy as ours. An appropriately designed theory describing their deformation under an applied magnetic field predicts a prolate–oblate transition in that case[13].

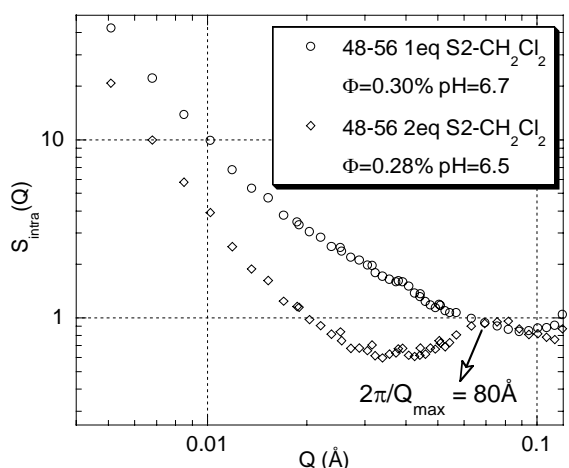

Fig. 3. SANS curves for the $PB_{48}$-$PGA_{56}$ copolymer mixed with 1 and 2 equiv. of the hydrophobic nanoparticles S2-$CH_2Cl_2$. In the small-Q region, the form factor of the aggregates follows a power law with slope approximately –2, which is characteristic of a flat sample. The structure peak due to particles at contact is absent for 1 equivalent.

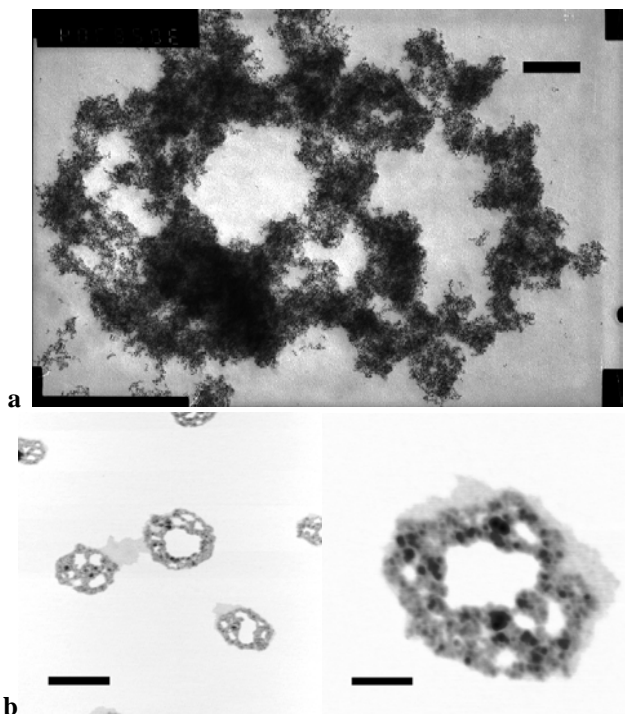

Fig. 4a) TEM picture of a magnetic membrane made of $PB_{48}$-$PGA_{56}$ and 1 equiv. of S2-$CH_2Cl_2$, showing only the nanoparticles because of the low electron density of the polymer compared to the iron oxide (bar length is 333nm);

b) AFM pictures of the same sample (the bar measures 500nm on the left and 150nm on the right). The uniform film visible around the objects is likely the pure copolymer bilayer spreading onto the mica substrate, with a measured height $\delta_{AFM}$=4.5nm. The vesicles appear holey on the pictures because the strong adhesion of those soft shells on high energy surfaces (freshly cleaved mica for AFM or Formvar coated graphite for TEM) necessitates the ripping of the bilayers (a sort of like pealed orange skins!).

(C). Vesicles self-assembling in water from $PB_{48}$-$PGA_{56}$ can also be filled in their inner aqueous compartment with a neutral ferrofluid just as S1S-$Na_3$Cit, thereby obtaining magnetic polymersomes analogous to magnetoliposomes[14], which are lipid vesicles loaded with an aqueous ferrofluid at pH7. In the same way as for magnetic liposomes, after preparing the vesicles one can separate them from non encapsulated nanoparticles by column chromatography with Sephacryl S1000 as separating medium[14]. The sorting process is studied by a look at the structure factor $S_{intra}(Q)$.

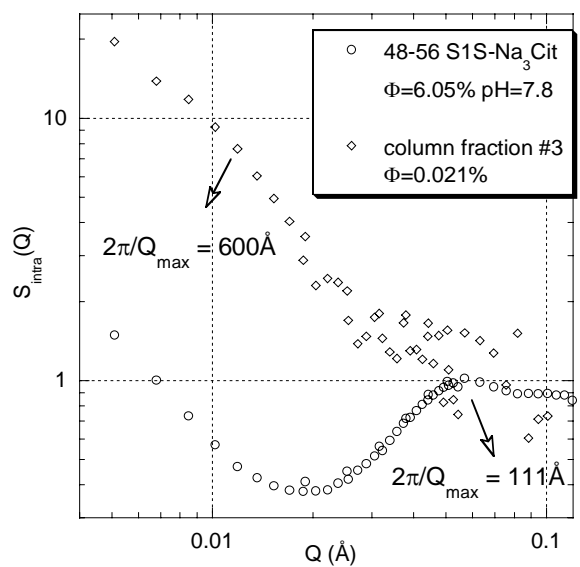

Fig. 5. SANS curves for $PB_{48}$-$PGA_{56}$ vesicles prepared in the citrated ferrofluid S1S-$Na_3$Cit before and after column fractionation. The low-Q region is a part of the form factor of large 3-d objects present in both samples. The structure peak in the initial unsorted dispersion gives a local concentration $\Phi_{local}$=11% inside the vesicles superior to the average concentration $\Phi$=6% (from iron titration). The less pronounced shoulder still visible on the curve after passage through the column corresponds to $\Phi_{local}$=0.07%, which is 3–4 times higher than the average $\Phi$=0.02%. The enrichment of loaded vesicles compared to free particles is thus concomitant with a global dilution of all the objects and a partial leak-out of the vesicles, maybe due to a larger size of copolymer vesicles than liposomes described in [14], compared to the porosity of the separating medium.

(D). Solutions of linear PAM with 2 vol. equivalents of S1S-$Na_3$Cit citrated ferrofluid at low and high [$Na_3$Cit]

We finish this article by presenting another example of polymer–ferrofluid hybrid leading this time to 1-d objects.

For this purpose, we dissolved long linear chains of PAM 5–6x10$^6$g/mol having a (calculated) radius of gyration in water around 100nm. We mixed the solutions with the S1S-Na$_3$Cit ferrofluid at two final values of the unbound tri-sodium citrate salt concentration [Na$_3$Cit]=8mM and 50mM respectively. The volume ratio of nanoparticles relatively to polymer was kept at a constant value (=2).

Table 2. Intrinsic viscosity [η] (g$^{-1}$.mL) measured by capillary viscosimetry for PAM solutions mixed with 2 vol. of S1S-Na$_3$Cit citrated ferrofluid at two values of [Na$_3$Cit]:

|  | 50mM citrate | 8mM citrate |
|---|---|---|
| PAM only | 535 | 595 |
| PAM + particles | 495 | 435 |

These data above show that the conformation of chains with and without nanoparticles does not vary a lot at high citrate concentration (8% decrease at 50mM free citrate), whereas the presence of nanoparticles decreases the chain swelling significantly when the citrate concentration is lower (27% decrease at 8mM citrate). The same effect of lowering the citrate concentration on the shrinking of the polymer chains by the nanoparticles is also observed by dynamical light scattering: the hydrodynamic diameter of the polymer-ferrofluid complexes varies from $d_H$=107nm at 50mM of unbound citrate down to $d_H$=78nm at 8mM. Therefore we can conclude that the nanoparticles strongly interact with the long linear polyacrylamide chains and that this coupling becomes stronger at lower unbound citrate salt concentration in equlibrium with the citrate ligands. An adsorption of the particles onto the chains is thus very probable, because it explains the specific effect of the citrate concentration by a competition between the citrate ligands and the polar amide groups of the polymer to access the surface of iron oxide. In the case of complete adsorption, we calculate that on average 10$^2$ particles adsorb on a single chain. Having shown that the polymer–particles complexes behave as microgels swelling and shrinking in a similar way as cross-linked ferrogels[15], we have a look at their morphology.

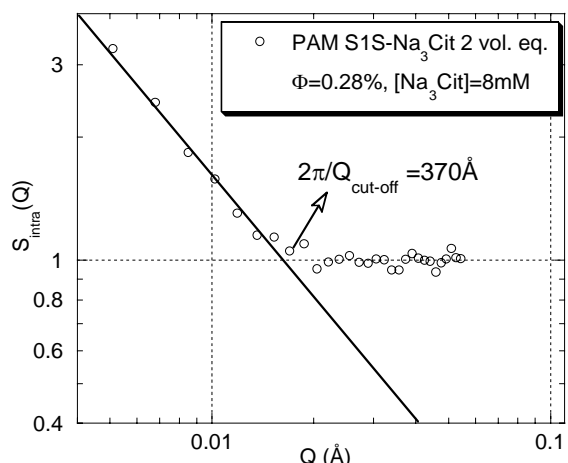

Fig. 6. SANS curve for the PAM solution mixed with 2 vol. of S1S-Na$_3$Cit at low unbound citrate concentration 8mM. The power law of slope –1 is the form factor of rigid rods, which diameter is estimated about 37nm from the cut-off.

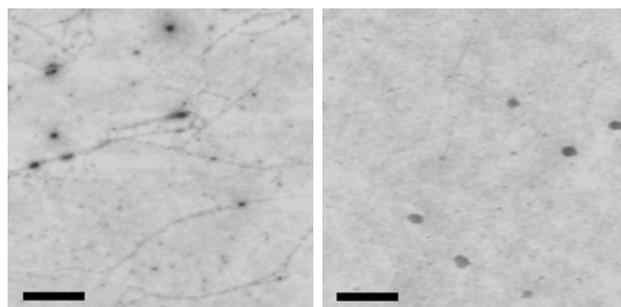

Fig. 7. AFM pictures of PAM solutions in S1S-Na$_3$Cit on mica at 8mM (left) and 50mM citrate (right, bar length is 500nm). We see necklaces consisting of spheres about 80nm in diameter linked to rigid segments up to 1µm long.

Now that we elucidated the self-assembly of those 1-d, 2-d or 3-d hybrid objects made of magnetic nanoparticles and (co)polymers, the next steps of this study will consist in attempting to modify their shape thanks to a magnetic field. Their magnetic response could be studied by anisotropic SANS measurements. In the case of a static B field first, one easily predicts to obtain a simple ellipsoidal deformation. Then with a time varying field B (for instance a rotating field) we could generate further complex shapes like the starfish-like ones well known for concentrated ferrofluid droplets. Preparing those original shapes with samples at the mesoscopic scale represents a challenge for the physical chemistry community. The aggregates of ferrofluids with di-block rod-coil copolymers or linear homopolymers open a possible route to reach this goal, also offering the possibility to freeze the structures by cross-linking unsaturated blocks like polybutadiene.